%% file: paper.tex
\documentclass[sigconf]{acmart}

\acmConference[P-RECS'19]{2nd International Workshop on Practical Reproducible Evaluation of Computer Systems}{June 24, 2019}{Phoenix, AZ}

\usepackage{balance}

\begin{document}

\title[Applicability Study of PRIMAD to LIGO gravitational wave search workflows]{Applicability study of the PRIMAD model to LIGO gravitational wave search workflows}

\author{Dylan Chapp}
\email{dchapp@udel.edu}
\affiliation{
  \institution{University of Delaware}
}

\author{Danny Rorabaugh}
\email{dror@utk.edu}
\affiliation{
  \institution{University of Tennessee, Knoxville}
}

\author{Duncan A. Brown}
\email{dabrown@syr.edu}
\affiliation{
    \institution{Syracuse University}
}

\author{Ewa Deelman}
\email{deelman@isi.edu}
\affiliation{
  \institution{University of Southern California}
}

\author{Karan Vahi}
\email{vahi@isi.edu}
\affiliation{
  \institution{University of Southern California}
}

\author{Von Welch}
\email{vwelch@iu.edu}
\affiliation{
    \institution{Indiana University, Bloomington}
}

\author{Michela Taufer}
\email{taufer@gmail.com}
\affiliation{
  \institution{University of Tennessee, Knoxville}
}

\renewcommand{\shortauthors}{}

\input{abstract.tex}


\maketitle

\input{section_1.tex}

\input{section_2.tex}

\input{section_3.tex}

\input{section_4.tex}

\input{section_5.tex}

\input{acknowledgements.tex}

\balance

\bibliographystyle{ACM-Reference-Format}
\bibliography{bibliography} 

\end{document}

%% file: abstract.tex
\begin{abstract}
The PRIMAD model with its six components (i.e., Platform, Research Objective, Implementation, Methods, Actors, and Data), provides an abstract taxonomy to represent computational experiments and enforce reproducibility by design. 
In this paper, we assess the model applicability to a set of Laser Interferometer Gravitational-Wave Observatory (LIGO)  workflows from literature sources (i.e., published papers). Our work outlines potentials and limits of the model in terms of its abstraction levels and application process.
\end{abstract}

%% file: section_1.tex
\section{Introduction}
The ability of the scientific community to incrementally build on experimental 
results depends strongly on the ability to trust that those results are not 
accidental or transient, but rather that they can be reproduced to an acceptably 
high degree of similarity by subsequent experiments. This notion of 
reproducibility is magnified both in importance and difficulty in the setting of 
computational science workflows~\cite{Stodden2016,Stodden2018}.
An increasingly large fraction of scientific 
endeavors depend on or incorporate computational elements, which in turn opens 
the door to reproducibility challenges associated with the implementation of 
those computational elements. 

In order to reason about  and assess reproducibility in the computational 
context, the PRIMAD model~\cite{Freire2016} was proposed. PRIMAD breaks 
reproducibility into six named components (i.e., Platform, Research objective,
Implementation, Methods, Actors, and Data), each of which represents an element 
of a computational experiment where reproducibility can be enforced by design, 
or conversely where a lack of such design can allow irreproducibility to seep in 
and corrode the overall integrity of the experiment.  

To evaluate the efficacy of PRIMAD as a tool for characterizing the 
reproducibility of real-world computational science workflows, we examine 
computational workflows used to detect gravitational waves using data from
the Laser Interferometer Gravitational-Wave Observatory 
(LIGO)~\cite{TheLIGOScientific:2014jea} and the Virgo 
Observatory~\cite{TheVirgo:2014hva}. These 
computational workflows are designed to detect various 
astronomical events, including binary black hole 
mergers~\cite{Abbott:2016blz,Abbott:2016nmj,Abbott:2017vtc,Abbott:2017gyy,Abbott:2017oio}
and binary neutron star mergers~\cite{TheLIGOScientific:2017qsa}.

There are three factors that make the gravitational-wave search workflows particularly appropriate 
for our post-hoc study through the lens of PRIMAD: (1) LIGO and Virgo have 
reached a mature status with findings that have been recognized by the 
scientific community at large; (2) gravitational-wave search workflows support high impact scientific 
findings (i.e., empirical confirmation of the existence of gravitational waves) 
that are subject to the highest levels of scrutiny from the broader scientific 
community; and (3) the workflows consume a large amount of data and have high 
internal complexity, leading to reproducibility challenges in terms of the 
implementation and data management components of PRIMAD.

We tackle the study of gravitational-wave search workflows in a post-hoc fashion using literature 
sources (i.e., published papers), rather than from the runtime point of view. 
The study mimics the effect of scientists in replicating 
the work done by others  as a starting point for new research. Specifically, we 
 study two gravitational-wave searches presented by the LIGO Scientific Collaboration and the Virgo Collaboration, and a third search presented by an independent group of scientists\footnote{We note that this independent group contained former LIGO Collaboration members. However, they only had access to publicly available data, code, and information in their analysis.}. These are: (1) the O1 Binary Catalogue~\cite{Abbott2016}; (2) the O2 Binary 
Catalogue~\cite{Abbott2018}; and (3) the Open Gravitational Wave Catalogue 
(1-OGC)~\cite{Nitz2019}. The findings are reported in this paper and organized 
in terms of the six components of PRIMAD.

%% file: section_2.tex
\section{The PRIMAD Model}
Large-scale scientific applications are composed of complex workflows, which in
turn are composed of and execute a series of experimental, computational, and 
data manipulation steps in one or multiple scientific domains, with at least one 
computational element. Research teams executes such workflows to gain insights 
that (1) confirm or disprove a hypothesis or (2) discover novel behavior or 
phenomena. 
 
The PRIMAD Model~\cite{Freire2016} is a powerful method to describe these 
workflows in terms of this 6-component space:
\begin{align*}
    \text{P} &= \text{Platform / execution environment / context} \\
    \text{R} &= \text{Research objectives / goals} \\
    \text{I} &= \text{Implementation / code / source-code} \\
    \text{M} &= \text{Methods / algorithms} \\
    \text{A} &= \text{Actors / persons} \\
    \text{D} &= \text{Data} 
\end{align*}
 
The model integrates multiple degrees of abstraction and is moldable to fit the 
characteristics of different scientific studies across domains. 

A \textit{platform} can be a simple execution environment such as a virtual 
machine or container; a more sophisticated combination of high-end clusters for 
computing and data processing; or a combination of remote sensing devices 
plugged into systems for collecting data, processing the data in preparation for 
analysis, analyzing the data, and displaying the results possibly through data 
visualization. 

A \textit{research objective} includes a specification of acceptable result 
variability provided by the scientist. If the scientist requires bitwise 
identical results (e.g., for regulatory reasons or out of technical necessity as 
in some techniques from mathematical physics and experimental 
mathematics~\cite{Bailey2012}), then this specification must be incorporated 
into the research objectives. Alternatively, if the scientist requires looser 
tolerances (e.g., convergence to a certain energy threshold as in ensemble 
simulations of protein folding trajectories~\cite{Piana2014}), this specification 
should be expressed as part of the research objective. 

The implementation can comprise a single executable or source-code (e.g., from 
benchmark applications such as the CORAL procurement benchmarks~\cite{Wu2015}, 
to production-grade applications such as ALE3D~\cite{Mcclelland2014}), or 
multiple software artifacts (e.g., solver packages such as 
HYPRE~\cite{Falgout2002}) that support one or more research objectives. 

Methods are broadly defined by individual computational algorithms, heuristic 
techniques, step-by-step procedures, or combinations thereof. 

Actors are those who perform the scientific study (e.g., scientists, users, 
technicians). They may have designed the overall workflow or individual parts of 
it, or simply use it. 

Data refers to all the data that are ingested by the initial stage of a workflow 
and those data objects that intermediate stages of the workflow generate.

%% file: section_3.tex
\section{Theoretical Scenarios Exhibiting Reproducibility Concepts}

Changes in the six components of the model generate different scenarios that 
exhibit different reproducibility behaviors~\cite{Claerbout1992,Donoho2010}. We 
envision six scenarios of interest (i.e., scenarios that are plausible for 
real-world scientific workflows). These scenarios are  summarized
in Table~\ref{tab:primad_scenarios}. Note that this table is not meant to be an 
exhaustive list.

\begin{table}[!htb]
    \centering
    \includegraphics[width=\columnwidth]{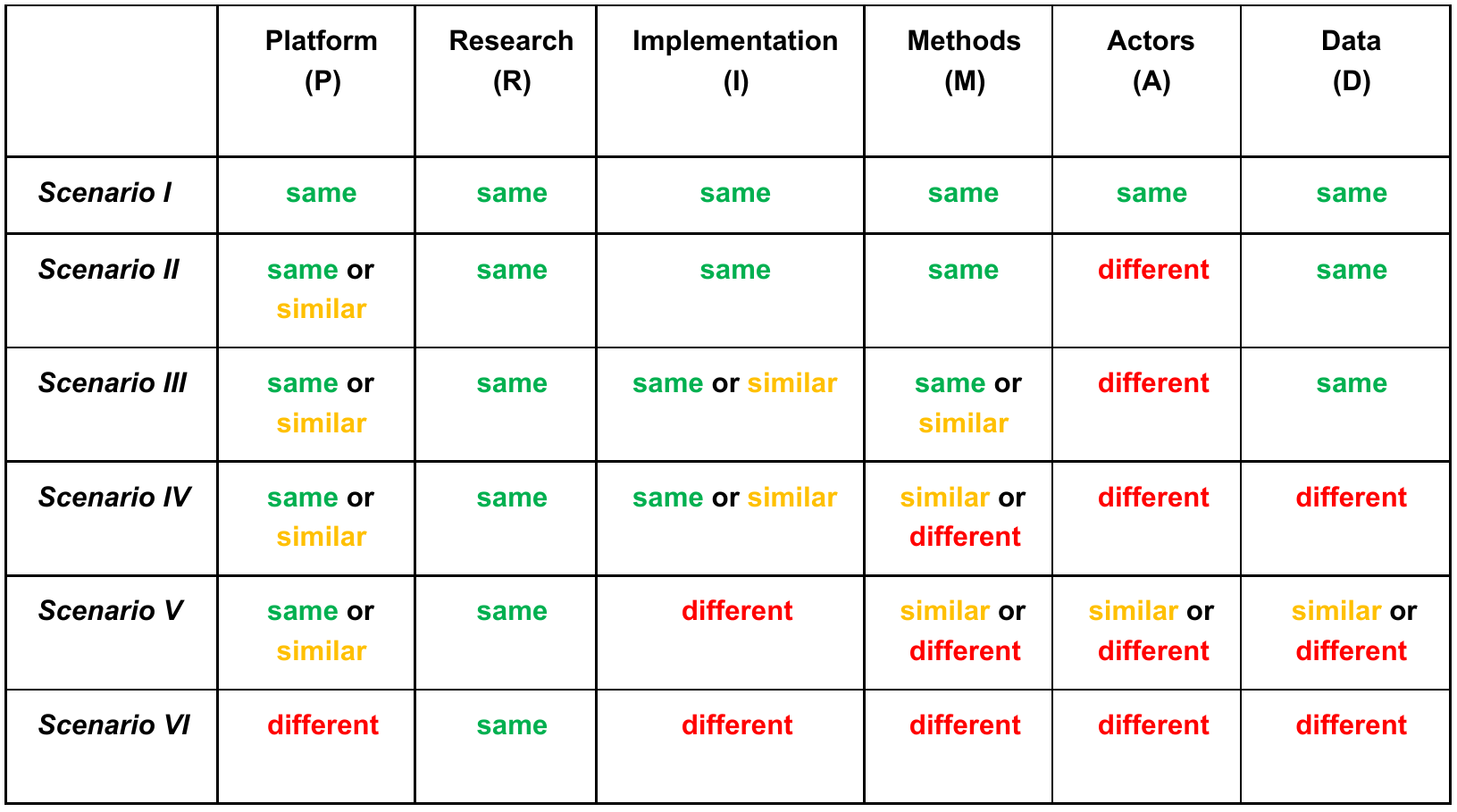}
    \caption{Overview of six scenarios exhibiting relevant reproducibility 
    outcomes. Concepts like similar mean some overlapping while different 
    implies complete disjoint component features or items.}
    \label{tab:primad_scenarios}
\end{table}

The first scenario is one in which the six PRIMAD components are all the same. 
In other words, the workflow is replayed using the same data and the same 
computational (software and hardware) environment. The same team of actors is 
able to produce results that confirm or disprove the hypothesis 
previously tested with the same workflow. Actors can repeat a measure over and 
over again, within the expected precision or threshold of acceptable 
variability. Challenges associated to this scenario are: (1) the platform 
availability can be an issue when dealing with workflows running on, for example, 
leadership class computing clusters; (2) documentation deficiencies can be an 
issue when dealing with complex workflows, for example, those concerning 
multiscale modeling simulations; and (3) run-to-run variability can occur, even 
when all six factors of the PRIMAD model are held fixed, due to hardware 
characteristics, platform concurrency, nondeterministic application behavior,
and other factors. Specifically, aspects of the chosen methods or implementation 
may be inherently nondeterministic, as in the case of dynamic 
multithreading~\cite{Utterback2017}.
 
The second scenario is one whose actors change, (e.g., a different team in the 
same or a different institution) and while the platform may be different, such 
differences shall be limited (e.g., different operating system version, upgraded 
library). In other words, by using the same data and the same or similar 
computational (software and hardware) platform, a different team of actors is 
able to produce results that confirm or disprove the hypothesis previously 
tested with the same workflow. In addition to the challenges of Scenario 1, this 
scenario is vulnerable to improper documentation dissemination, for example, 
sharing documentation with incomplete or ambiguous descriptions. This scenario 
and its challenges match such efforts as the Student Cluster Competition~\cite{Garrett2018}. 
 
The third scenario explores the impact of varying, to different degrees, all 
PRIMAD components of the model except the research objectives and the data. By 
using the same data (e.g., the same initial conditions such as pressure, 
temperature, and volume in a molecular dynamic simulation) but allowing limited 
variability of implementations and platforms or both, along with published 
material and related artifacts, an independent team is able to produce result 
that confirm or disprove the same hypothesis presented in the published 
material. The challenges associated with this, and following scenarios, depend
on the degree of variation in the individual components of the PRIMAD model 
that we allow.
 
The fourth scenario is a scenario that is less constrained than the previous 
one. It explores the impact of varying all PRIMAD components except the research 
objectives. This scenario incorporates the additional challenge of collecting 
new data according to the data generation description in one or multiple 
original publications. The study can be performed by using the same or similar 
implementations and platforms. In other words, minor variations in 
implementation and platform are acceptable in this scenario. Along with the 
published material and related artifacts, an independent team is able to produce 
results that confirm or disprove the same hypothesis.
 
The fifth scenario studies the ability of a team of actors to reproduce the 
research of another team using a different implementation (e.g., in molecular 
dynamics simulations using AMBER rather than CHARMM). They address the same 
research objectives, but the level of variability in workflow outcomes depends 
upon the changes in implementation.
 
The sixth scenario tackles the study of a scientific application for which the 
workflow has a substantial change of the platform and implementation. 
Specifically, this scenario deals with a major technological transformation 
(e.g., from multiprocessor to accelerated platforms). All components of the 
model space in which the workflow is executed, with the exception of the 
research objective, change sufficiently and thus these variations cannot be 
ignored.

%% file: section_4.tex
\section{Empirical Scenarios Exhibiting Reproducibility Concepts in the LIGO Project}

Our post-hoc study of the PRIMAD applicability to gravitational-wave search workflows falls into Scenario V. 
When we use the term workflow, we refer to the analysis of a specific set of data using a specific search pipeline. In practice, this may involve running several identically-configured workflows on smaller data sets to produce a single search result. 
Specifically, we target the gravitational wave search in three literature sources: (1) the LIGO-Virgo O1 Binary Catalogue~\cite{Abbott2016}; (2) the LIGO-Virgo O2 Binary Catalogue GWTC-1~\cite{Abbott2018}; and (3) the Open Gravitational Wave Catalogue (1-OGC)~\cite{Nitz2019}, produced independently of the LIGO-Virgo Collaborations. 
The two LIGO-Virgo Binary Catalogue papers~\cite{Abbott2016,Abbott2018} describe two matched filtering workflows built on two implementations or source codes (with PyCBC~\cite{Usman:2015kfa,TheLIGOScientific:2016qqj,Nitz:2017svb,Nitz2016} and GstLAL~\cite{TheLIGOScientific:2016qqj,Messick2017,Mukherjee:2018yra,Sachdev:2019vvd} data analysis suites, respectively). 
The LIGO-Virgo O1 Binary Catalogue paper~\cite{Abbott2016} and its constituent workflows use the dataset from Observing Run 1 of the LIGO detectors (O1). 
The LIGO-Virgo O2 Binary Catalogue paper~\cite{Abbott2018} and its constituent analyses use the dataset from Observing Run 2 of the LIGO detectors (O2) as well as data from the VIRGO detector. 
The 1-OGC paper~\cite{Nitz2019,Nitz2019a} describes a run of the PyCBC matched filtering workflow on the public LIGO O1 data using an updated pipeline configuration compared to the analysis in the LIGO-Virgo O1 Binary Catalogue. 
The relationships between the workflows and their parent publications are summarized in the first two rows of Table~\ref{tab:ligo_primad_table} together with the six components of the PRIMAD model.

\begin{table}[!htb]
    \includegraphics[width=\columnwidth]{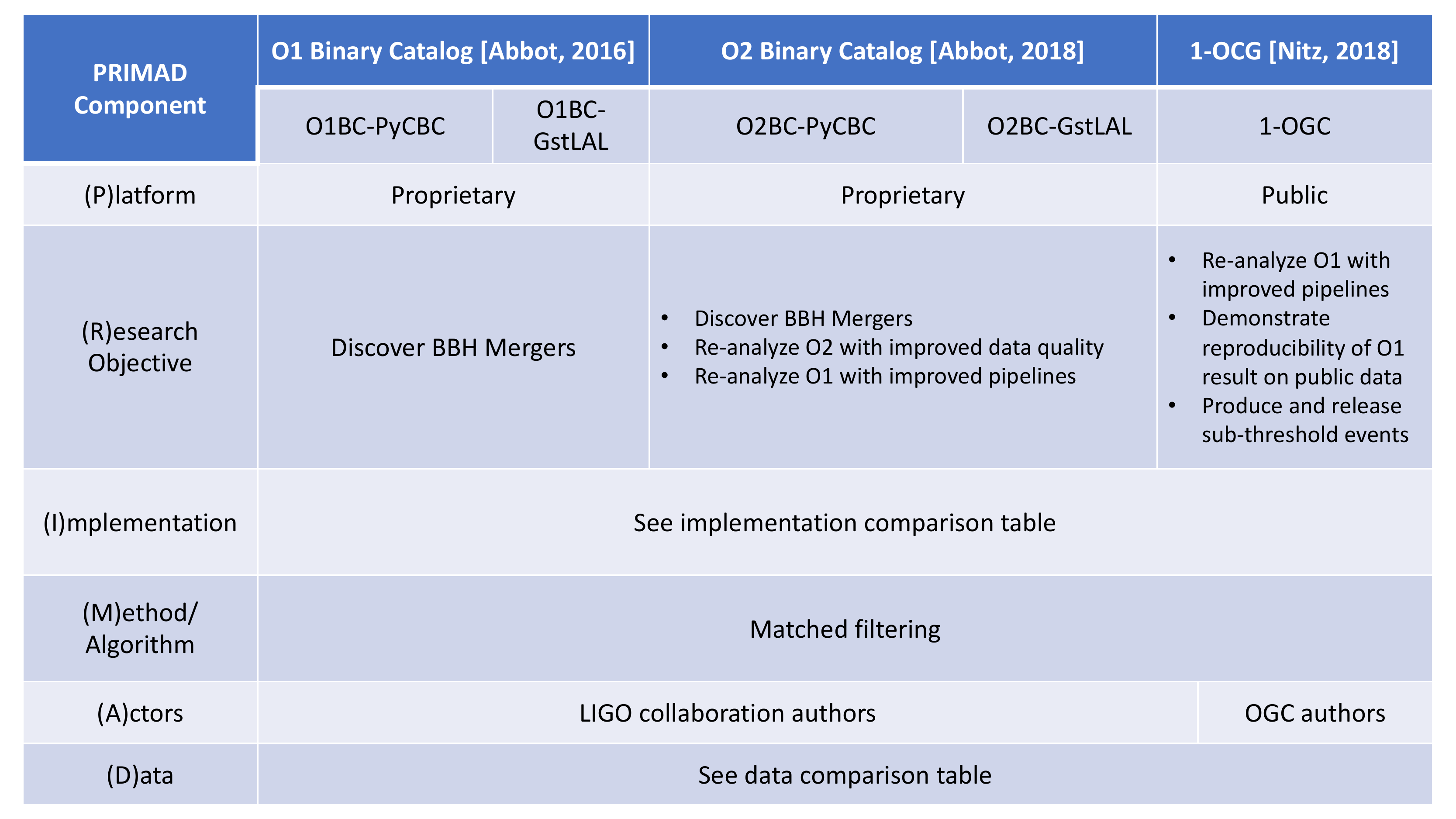}
    \caption{ Overview of LIGO analyses considered in our case study in terms of 
    the 6 PRIMAD model components.}
    \label{tab:ligo_primad_table}
\end{table}

\begin{table}[!htb]
    \includegraphics[width=\columnwidth]{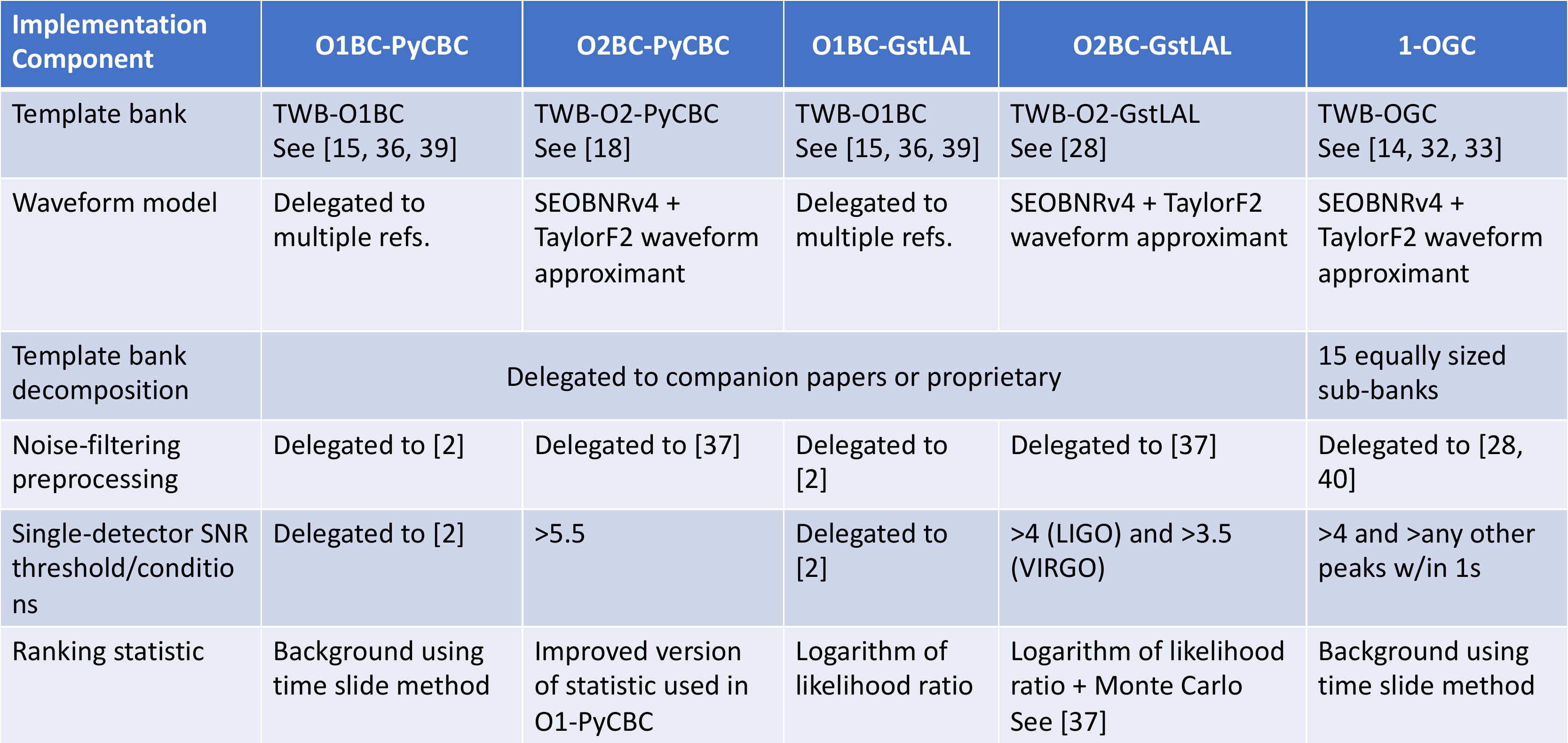}
    \caption{Matched filtering implementation across the LIGO data workflows 
    targeted in our study  (i.e., O1BC-PyCBC, O2BC-PyCBC, O1BC-GstLAL, 
    O2BC-GstLAL, 1-OGC).}
    \label{tab:implementation_sub_table}
\end{table}

\begin{table}[!htb]
    \includegraphics[width=\columnwidth]{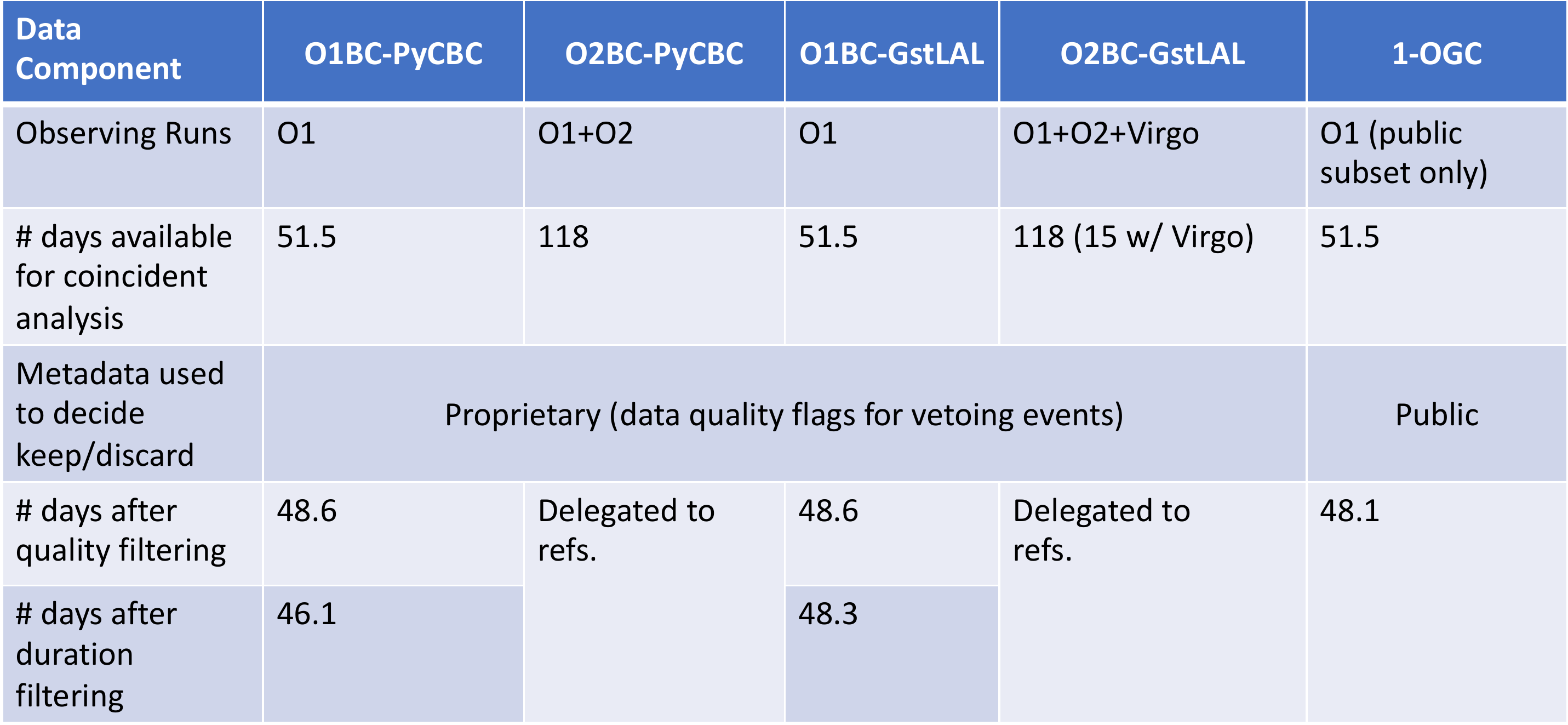}
    \caption{Data details across the targeted LIGO workflows targeted in our 
    study  (i.e., O1BC-PyCBC, O2BC-PyCBC, O1BC-GstLAL, O2BC-GstLAL, 1-OGC).}
    \label{tab:data_sub_table}
\end{table}

In the table, the artifacts are executed on proprietary or public \textit{platforms}. We consider proprietary platforms to be those platforms that are only the actors of the workflow have access to. In contrast we consider platforms to be public when they are accessible to a broader set of potential actors (e.g., the Open Science Grid (OSG)~\cite{Pordes:2007zzb,Sfiligoi:2010zz}). 

The research objective of the workflows is to detect binary mergers using the data gathered by the LIGO and Virgo detectors. The workflows may also re-analyze previous data with improved methods and tuning to re-assess the significance of previous events, or discover previously undetected events; this is considered similar enough to the core research objective to be considered the same within the context of PRIMAD.

The descriptions of matched filtering workflows provide numerous details in terms of implementation choices (e.g., choices of signal-to-noise ratio (SNR) thresholds for detecting significant matches) and data management choices (e.g., choices for how detector instrument noise is filtered and mitigated). The abundance of details concerning implementation and data variations across the matched filtering studies are summarized in Table~\ref{tab:implementation_sub_table} and Table~\ref{tab:data_sub_table} respectively. 

Descriptions of the search pipeline implementation are listed in the leftmost column of Table~\ref{tab:implementation_sub_table}. 
These sub-components define the particular way that a workflow (e.g., the O1BC-PyCBC workflow) implements matched filtering (i.e., the method component). 
Specifically, we have highlighted the following implementation sub-components: the template bank used, the waveform model used to generate the template bank, the policy used to decompose the template bank into sub-banks, any additional noise reduction or filtering steps performed, the signal-to-noise ratio threshold for selecting candidate matches (i.e., events that could potentially be binary black holes), and the ranking statistic used rank candidate events. 

For the template bank component, we abbreviate the banks used for each workflow.
TWB-OGC refers to the template bank used in the 1-OGC search workflows, the properties of which are defined in~\cite{brown2012detecting,owen1996search,owen1999matched}.
TWB-O1BC refers to the common template bank used in both the O1BC-PyCBC and O1BC-GstLAL workflows, the properties of which are defined in~\cite{taracchini2014effective,purrer2016frequency,capano2016implementing}.
TWB-O2BC-PyCBC refers to the template bank used for the O2 Binary Catalogue PyCBC workflow, the properties of which are defined in~\cite{DalCanton2017}.
TWB-O2BC-GstLAL refers to the template bank used for the O2 Binary Catalogue GstLAL workflow, the properties of which are defined in~\cite{Mukherjee:2018yra}.

We find that for some components of the workflow, details are predominantly delegated to one or more cited publications. In other cases, configuration of the workflow is only available by examining the workflow's source code or the configuration files used in a specific analysis. We also find that some workflow configuration files are public (e.g. \cite{PyCBC-Config,Nitz2019a}), but others require proprietary access. Even for configuration and software repositories that are public, it can be difficult to determine which versions were used for a particular analysis.
Similarly for Table~\ref{tab:data_sub_table}, we describe sub-components of the Data PRIMAD component in the leftmost column. 
The sub-components we describe include: the source of the raw input data (i.e., which observing runs the data comes from), the number of coincident days of data available (i.e., data from when both LIGO detectors were operating), the choice of metadata used to filter data prior to the matched filtering search, the number of days of observation data  remaining after filtering for data quality, and the number of days of observation data remaining after filtering insufficiently long-lasting events (i.e., filtering based on event duration). 
These details are largely provided across the chosen workflows, unless those details are proprietary. 

Across all of the workflows described in the LIGO papers used for our study, the high-level method that is implemented is matched filtering~\cite{Allen:2005fk}, that is, the procedure of matching a stream of data from the LIGO detectors against template waveforms derived from numerical general relativity. Matches against these templates may indicate the presence of astronomical events such as binary black hole mergers, and thus are evaluated for statistical significance.

%% file: section_5.tex
\section{Discussion and Conclusions}
We conclude with a reflection on PRIMAD, beginning with general benefits and challenges of the model itself. 
Then we evaluate our post-hoc application of PRIMAD to the LIGO workflows, which informs best practices for future runtime applications of PRIMAD in the execution of scientific workflows.

PRIMAD is a general model to guide reproducibility. 
It helps meet an acute need in the scientific community to ground reproducibility, yet it is inherently abstract due to its applicability across all scientific domains, leading to challenges in establishing a useful level of specificity. 
When researchers want to share their findings,  many research teams (or \textit{actors}) make best efforts at including all relevant information to enable reproducibility.
Without standardization, the decisions about what constitutes ``relevant information'' are inevitably ad-hoc, and may not be uniform from publication to publication or across multiple workflows within a single publication.
Thus, PRIMAD offers a framework with which to build sustainable reproducibility in an uniform fashion across scientific domains.

Challenges arise in identifying how specific the components of PRIMAD need to be defined in order to guarantee both consistent applicability across workflows in a specific domain and desired levels of reproducibility. 
For example, the division between implementation and methods is disputable. 
Minor adjustments to an algorithm would generally fall into implementation, yet it is hard to determine when changes are substantial enough to call it a new algorithm and thus a change in methods. 
In other cases, the effects of the actors on reproducibility may be difficult to document. 
Even within the same research group with consistent leadership, research objectives, and computational environments, changes in team members and shifts in member responsibility can introduce unacknowledged sources of variability. 
It appears to be a difficult problem to appropriately document the knowledge and experience that is applied to the elements of a workflow. 
Finally, different datasets may play different roles within a workflow. 
As a partial remedy, we propose a differentiation between input data (including initial parameters), intermediary data (which depends heavily on methodology and implementation), and output artifacts (having direct relevance to the research objective). 
In order to meet these challenges, we recommend that each field of science develop its own domain-appropriate refinement to PRIMAD.

The LIGO workflows demonstrate scientific research performed with a clear desire for reproducibility. 
Therefore, even though the authors do not apply an explicit reproducibility framework, we were able perform a post-hoc evaluation of the workflows with the PRIMAD model. 
In particular, we find that out of the six PRIMAD components, the implementation and data components are given the most attention in terms of documented details, as summarized in Table~\ref{tab:implementation_sub_table} and Table~\ref{tab:data_sub_table}. 
One PRIMAD component not described with the same level of details as the other components is the platform used in the workflows. 
This may be due to multiple factors such as: (1) use of proprietary platforms; (2) prioritization of implementation or data management details over platform details; and (3) the technical challenge of tracking all the platform components on which a workflow manager deploys jobs. 
Uncertainty in the reason for incomplete information is one of a few limitations we found in post-hoc application of PRIMAD. 
Non-uniform descriptions of PRIMAD components and the necessity of citation chasing to fill in details also arise, though are mitigated by documentation efforts of actors within a study and overlap of actors between studies.

Challenges listed above can be mitigated by moving from a post-hoc to a runtime application of PRIMAD from the early stages of scientific experimentation. 
The PRIMAD components should be periodically referred to as a checklist for assessing the reproducibility of workflows, both as they are developed and during the transition from experimentation to publication. 
Moreover, rather than hampering scientific progress by imposing requirements inconsistent with a particular scientific workflow, we demonstrated with the various scenarios in Section 4 how such a framework can be a guide to ensure the desired level of reproducibility.

A wide adoption of PRIMAD would be most achievable through involvement of all contributors to the scientific process. 
That is, tools that researchers use to do computational experiments must: (1) be aware of the PRIMAD components; and (2) provide users with access to details about their experiments, organized in terms of the PRIMAD components, so that those details can be disseminated in publications. 
In particular, we look at the role of workflow managers in the computational components of a scientific workflow. 
In the case of the PyCBC matched filtering pipelines, the actors used Pegasus~\cite{Deelman2005,Brown2007} to distribute constituent jobs to participating computing centers~\cite{Weitzel2017}. 
In the long run, workflow managers such as Pegasus can play a key role in tracking, for example, which computational platforms their jobs run on so that future research teams can attempt to reproduce results on similar platforms, and therefore automatically, deploying the PRIMAD model at runtime.

As part of our study of gravitational-wave search pipelines, both from the LIGO collaboration and from others,  we have identified the various implementation and data components that need to be identified in order for an outside actor to reproduce the results of these pipelines. 
We feel this axis of classification can serve as a foundation for describing all (or a large number of) gravitational-wave search pipelines and lowering the barriers for reproducing the results outside of the LIGO collaboration.

%% file: acknowledgements.tex
\begin{acks}
The authors thank Almadena Chtchelkanova and the U.S. National Science Foundation for
sponsoring the July 25, 2017 workshop that provided the foundation on the PRIMAD model.  The authors
also thank the workshop attendees: Michael A. Heroux, Sandia National Laboratories; Lorena A. Barba, Ronald Boisvert, National Institute of Standards and
Technology; Bruce Childers, University of Pittsburgh;
Juliana Freire, New York University; Carlos Maltzahn, University of California, Santa Cruz; Wilf Pinfold, Urban.Systems; Jeff Spies, Center for Open Science.
The work was supported by NSF awards OAC-1841399, OAC-1823405, OAC-1823378, and CCF-1841552.
\end{acks}